# The Role of CP violation in $D^0\overline{D^0}$ Mixing

G. BLAYLOCK, A. SEIDEN

*Santa Cruz Institute for Particle Physics*
*University of California at Santa Cruz*
*Santa Cruz, CA 95064*

and

Y. NIR

*Dept. of Particle Physics*
*Weizmann Institute of Science*
*Rehovot 76100, Israel*

## ABSTRACT

In current searches for $D^0\overline{D^0}$ mixing, the time evolution of "wrong-sign" decays is used to distinguish between a potential mixing signal and the dominant background from doubly-Cabibbo-suppressed decays. A term proportional to $\Delta M t$ in the expression for the time evolution is often neglected in theoretical discussions and experimental analyses of these processes. We emphasize that, in general, this term does not vanish even in the case of CP invariance. Furthermore, CP invariance is likely to be violated if the rate of $D^0\overline{D^0}$ mixing is close to the experimental bound. The consequence of either of these two facts is that the strongest existing measured bound is not applicable for constraining New Physics.

# 1. Introduction

The Standard Model predicts $D^0\overline{D^0}$ mixing that is orders of magnitude below the reach of present experiments. Consequently, a discovery of $D^0\overline{D^0}$ mixing would provide a clear signal of New Physics. Indeed, various extensions of the Standard Model allow the mixing rate to be close to the current experimental bound. Examples of such extensions are Supersymmetry with quark-squark alignment; additional sequential (namely, fourth generation) or nonsequential (e.g., left-handed singlet) up-type quarks; multiscalar models with or without natural flavor conservation; and leptoquarks. Upper bounds on $D^0\overline{D^0}$ mixing constrain the parameter space of these models.

In order to identify mixing experimentally, it is necessary to determine the charm flavor of neutral $D$ mesons at both their production and decay points. One common method is to use the decay $D^{*+} \to \pi^+ D^0$ (or $D^{*-} \to \pi^- \overline{D^0}$), where the charge of the pion determines whether a $D^0$ or a $\overline{D^0}$ was produced. A subsequent decay of the $D$ into a final state with the "wrong-sign" kaon (such as $D^0 \to K^+\pi^-$ or $\overline{D^0} \to K^-\pi^+$) provides a possible indication of mixing. However, for hadronic final states, wrong-sign decays can also come from direct decays via doubly-Cabibbo-suppressed (DCS) amplitudes. Although DCS rates are expected to be small (less than 1% of the Cabibbo-favored rates) they provide a substantial background for the rare process of mixing.

Traditionally, these two mechanisms for wrong-sign decays are distinguished by their different evolutions in time. However, there has been some confusion in the literature over the exact form of the time evolution of wrong-sign decays. An interference term that can potentially weaken the sensitivity to mixing is often neglected, with the justification that it vanishes in the case of CP invariance. This claim is incorrect. We will review the relevant formalism with an explanation of the confusion and will demonstrate, in addition, that the assumption of CP invariance is not likely to be valid in the charm system if the rate of $D^0\overline{D^0}$ mixing is close to the current experimental bound.

## 2. Time Evolution Formalism

We begin by reviewing the formalism of $D^0\overline{D^0}$ mixing, following the notation of reference 1. We define $p$ and $q$ as the charm eigenstate components in the mass eigenstates $|D_{1,2}\rangle$:

$$|D_1\rangle = p\,|D^0\rangle + q\,|\overline{D^0}\rangle,$$
$$|D_2\rangle = p\,|D^0\rangle - q\,|\overline{D^0}\rangle, \qquad (2.1)$$

with the normalization

$$|p|^2 + |q|^2 = 1. \qquad (2.2)$$

The two physical states evolve according to

$$|D_i(t)\rangle = e^{-iM_i t - \frac{1}{2}\Gamma_i t}\,|D_i(t=0)\rangle, \qquad (2.3)$$

with the physical masses and widths given by $M_i$ and $\Gamma_i$.

For the observation of mixing, we are interested in the evolution of the state $|D^0(t)\rangle$ that starts out as a pure $|D^0\rangle$ at $t=0$ and of the state $|\overline{D^0}(t)\rangle$ that is initially pure $|\overline{D^0}\rangle$. Using Eqs. (2.1) and (2.3), we arrive at

$$|D^0(t)\rangle = f_+(t)\,|D^0\rangle + \frac{q}{p}\,f_-(t)\,|\overline{D^0}\rangle,$$
$$|\overline{D^0}(t)\rangle = \frac{p}{q}\,f_-(t)\,|D^0\rangle + f_+(t)\,|\overline{D^0}\rangle, \qquad (2.4)$$

where

$$f_+(t) \equiv e^{-iMt - \frac{1}{2}\Gamma t}\,\cos\!\left(\tfrac{1}{2}\Delta M t - \tfrac{i}{4}\Delta\Gamma t\right),$$
$$f_-(t) \equiv e^{-iMt - \frac{1}{2}\Gamma t}\,i\sin\!\left(\tfrac{1}{2}\Delta M t - \tfrac{i}{4}\Delta\Gamma t\right), \qquad (2.5)$$

and

$$M \equiv \tfrac{1}{2}(M_1 + M_2), \qquad \Delta M \equiv M_2 - M_1,$$
$$\Gamma \equiv \tfrac{1}{2}(\Gamma_1 + \Gamma_2), \qquad \Delta\Gamma \equiv \Gamma_2 - \Gamma_1. \qquad (2.6)$$

We are primarily interested in the decay of these states into wrong-sign final states. Accordingly, we define four decay amplitudes:

$$A \equiv \langle f|H|D^0\rangle, \quad B \equiv \langle f|H|\overline{D^0}\rangle,$$
$$\overline{A} \equiv \langle \overline{f}|H|\overline{D^0}\rangle, \quad \overline{B} \equiv \langle \overline{f}|H|D^0\rangle, \tag{2.7}$$

which we intend to represent doubly-Cabibbo-suppressed amplitudes ($A$ and $\overline{A}$) and Cabibbo-favored amplitudes ($B$ and $\overline{B}$), with $\langle f|$ and $\langle \overline{f}|$ describing CP conjugate states. The decay amplitudes into wrong-sign final states are then

$$\langle f|H|D^0(t)\rangle = B\frac{q}{p}\left[\lambda f_+(t) + f_-(t)\right],$$
$$\langle \overline{f}|H|\overline{D^0}(t)\rangle = \overline{B}\frac{p}{q}\left[f_-(t) + \overline{\lambda}f_+(t)\right], \tag{2.8}$$

where

$$\lambda \equiv \frac{p}{q}\frac{A}{B}, \quad \overline{\lambda} \equiv \frac{q}{p}\frac{\overline{A}}{\overline{B}}. \tag{2.9}$$

One may note in passing that $\lambda$ and $\overline{\lambda}$ are independent of phase conventions even though $\frac{p}{q}$ and $\frac{A}{B}$ separately depend on those conventions.[2]

Under the assumption that $\Delta M \ll \Gamma$, $\Delta \Gamma \ll \Gamma$ and $|\lambda| \ll 1$ (as confirmed experimentally), but not necessarily that $\Delta \Gamma \ll \Delta M$ (though we will suggest later that this may also be a reasonable assumption if $\Delta M$ is close to the current experimental bound), the decay rates are given by

$$\Gamma[D^0(t) \to f] = \frac{e^{-\Gamma t}}{4}|B|^2\left|\frac{q}{p}\right|^2 \times$$
$$\left[4|\lambda|^2 + \left(\Delta M^2 + \frac{\Delta \Gamma^2}{4}\right)t^2 + 2\Re(\lambda)\Delta\Gamma t + 4\Im(\lambda)\Delta M t\right] \tag{2.10}$$

and

$$\Gamma[\overline{D^0}(t) \to \overline{f}] = \frac{e^{-\Gamma t}}{4}|\overline{B}|^2\left|\frac{p}{q}\right|^2 \times$$
$$\left[4|\overline{\lambda}|^2 + \left(\Delta M^2 + \frac{\Delta \Gamma^2}{4}\right)t^2 + 2\Re(\overline{\lambda})\Delta\Gamma t + 4\Im(\overline{\lambda})\Delta M t\right]. \tag{2.11}$$

In these expressions for the wrong-sign rates, the terms proportional to $|\lambda|^2$ and $|\overline{\lambda}|^2$ describe the contributions from DCS amplitudes, the terms proportional to $t^2$ describe the lowest order contributions from mixing, and the terms proportional to $t$ represent interference between mixing and DCS amplitudes.

## 3. Effects of CP Invariance

It is often stated[3] that the term in (2.10) that is proportional to $\Delta M t$ changes sign under CP, which implies that $\Im(\lambda) = 0$ if CP is conserved. This statement led the E691 experiment[4] to assume that this term could be neglected if CP violation is small, and moreover that it would average to zero when analysing a data set with equal numbers of $D^0$ and $\overline{D^0}$. As we will mention later, other arguments suggest that the interference term proportional to $\Delta \Gamma t$ is also small. Consequently, E691 quoted a primary result ignoring both terms and assuming no interference between mixing and DCS amplitudes. This assumption substantially improved the quoted sensitivity of the experiment.

In truth, CP invariance implies that $\Gamma[D^0(t) \to f] = \Gamma[\overline{D^0}(t) \to \overline{f}]$ which leads to the following three conditions:

$$\frac{|A|}{|\overline{A}|} = \frac{|B|}{|\overline{B}|} = 1,$$
$$\left|\frac{p}{q}\right| = 1, \quad (3.1)$$
$$\lambda = \overline{\lambda}.$$

However, CP does not relate $A$ to $B$. Therefore the relative strong phase of $A$ and $B$, and hence $\Im(\lambda)$, is not constrained. Dunietz[1] has previously emphasized this feature in the context of CP violation in the $B$ meson system. In fact, many authors[5] have considered a possible relative phase between $A$ and $B$ for $D$ decays into $K\pi$ final states.

A few general remarks can be made about the phase of $A/B$ that do not depend on specific models. The spectator diagrams that contribute to $D$ decay are shown in Figure 1. For the case where $|f\rangle$ is a multibody final state (e.g., $K^+\pi^-\pi^0$), these diagrams do not necessarily provide similar mixtures of resonant intermediate states (e.g., $K^{*+}\pi^-$ and $K^+\rho^-$). These intermediate states will provide nontrivial phases from Breit-Wigner propagators, which, in turn, lead to a phase in the ratio $A/B$.

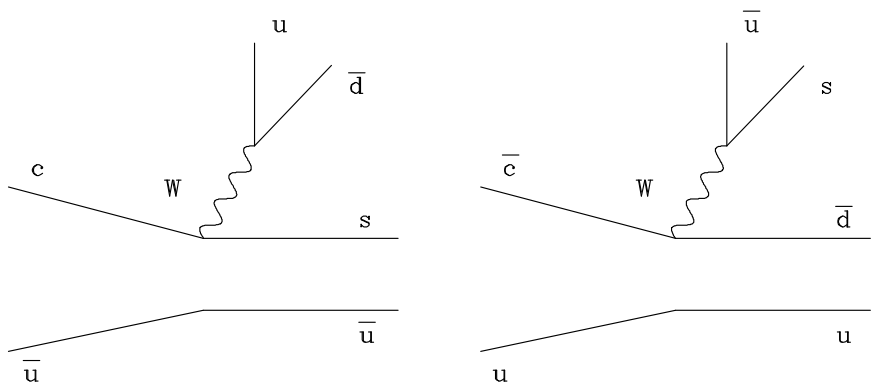

Figure 1: Cabibbo-favored (left) and DCS (right) decay diagrams.

The same observation applies even in two-body final states such as $D^0 \to K\pi$. In this case, all real intermediate states with the same strong quantum numbers (such as $K3\pi$) can contribute to $K\pi$ due to strong final state interactions. These can contribute differently to the two amplitudes $A$ and $B$, yielding non-zero phases in $A/B$. This is different from the case of $K$ decay to $2\pi$, where the only available channel of given isospin is the $2\pi$ channel, resulting in a uniquely determined strong phase.

The mistaken claim that the $\Delta M t$ term of (2.10) is odd under CP is partly a result of applying the fact that CP invariance is equivalent to T invariance, and noting that this term is odd in $t$. However, T reversal is more than just $t \to -t$. It also exchanges initial and final states, which can introduce new phases. The special case where the phase of $\lambda$ *can* be constrained is when the final state $\langle f|$ is a CP eigenstate. This is the main point of some B physics studies. In this case $A$ and $B$ refer to CP conjugate processes, and their phases are related by a CP transformation.

By itself, neglecting the $\Delta M t$ term in equation (2.10) will not have an important effect on fits to experimental data since another unknown term proportional to $\Delta \Gamma t$ remains. However, as we will discuss in the next section, there are arguments which suggest that $\Delta \Gamma$ may be small compared to $\Delta M$ in any New Physics models. This observation may have been what led the E691 experiment to neglect both interference terms when calculating their primary result. Because of the large correlations between terms in fits to experimental data, neglecting both interference terms leads to a much better apparent sensitivity to the remaining terms for mixing and DCS decays. Figure 2 indicates how these correlations come about in a hypothetical case where the interference contribution approximately cancels the contribution from pure mixing. This plot demonstrates that even when the time evolution deviates only slightly from the pure exponential form of DCS decays, a large contribution from mixing can be tolerated if it is offset by a destructive interference contribution. This implies that the fitted values for the interference contribution and the mixing contribution are strongly anti-correlated.

In addition to their primary result, E691 explored several specific interference terms which increased their quoted limits on mixing by up to a factor of 3.8. Preliminary studies from the E791 experiment[6] indicate sensitivities to mixing which are degraded by a factor of 3 to 5 when an arbitrary interference term is allowed.

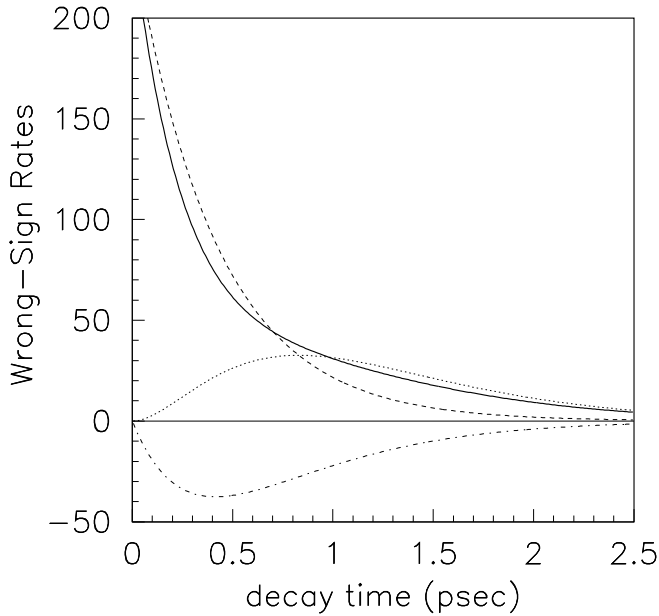

Figure 2: A hypothetical plot of the time dependence of wrong-sign decays. The dashed line represents the DCS contribution. The dotted line shows the contribution due to mixing. The dash-dot line shows the contribution from DCS-mixing interference when the interference is 30% of its maximum possible amplitude and destructive. The solid line is the sum of all three contributions.

## 4. Specific Models

The assumption made in various theoretical and experimental studies of $D^0\overline{D^0}$ mixing is that CP invariance is a good approximation for the relevant processes. In this section we will demonstrate that this assumption is, in general, unjustified; all reasonable models that allow a rate of $D^0\overline{D^0}$ mixing close to the experimental bound have new CP violating phases that affect the mixing. Consequently, if $\Delta M$ is close to the experimental bound, then the parameter $\lambda$ in Eq. (2.9) depends on both unknown strong phase shifts and unknown CP violating phases. On the other hand, *direct* CP violation is still likely to be negligible, which implies $|A/\overline{A}| = |B/\overline{B}| = 1$.

Another assumption, that $\Delta \Gamma \ll \Delta M$, is very likely to be a good approximation if $\Delta M$ is close to the experimental bound. The Standard Model contribution to $\Delta M$ is highly suppressed by small CKM angles; by the GIM mechanism, which is very significant for the intermediate down sector quarks; and by being fourth order in the weak coupling constant. New Physics, even at a scale much higher than the electroweak scale, could therefore easily dominate $\Delta M$. However, the Standard Model contribution to $\Delta\Gamma$ comes dominantly from tree level $W$-mediated decays. There is no current New Physics model that could enhance such decays by orders of magnitude.

The approximation $\Delta\Gamma \ll \Delta M$ leads, in turn, to the approximation $|q/p| \approx 1$, which simplifies the rate equations (2.10) and (2.11). It also makes clear that, contrary to previous suggestions[7], the interference term cannot be constrained using arguments about the sign of $\Delta\Gamma$. The approximation $\Delta\Gamma \ll \Delta M$ does not affect the experimental analysis, as the $\Delta\Gamma t$ and $\Delta M t$ terms cannot be distinguished.

We now consider various extensions of the Standard Model which allow large $\Delta M$ for the $D^0$ system[8]. For each of these models we consider the following questions:

(a) Is CP a good symmetry in the relevant processes?

(b) If $\Delta M$ is significantly enhanced, is $\Delta\Gamma$ similarly enhanced?

1. *Quark-Squark Alignment*

In supersymmetric models with quark-squark alignment[9], there are new contributions to $\Delta M$ from box diagrams with gluinos and squarks. The gluino-quark-squark coupling can be estimated, $g_{\tilde{g}\tilde{u}c} \sim g_s \sin\theta_c$, leading to the conclusion that the new contribution is of the order of the experimental bounds. (This model is unique in that it can be excluded by improved bounds on $D^0\overline{D^0}$ mixing.) There are arbitrary new CP violating phases in the gluino mixing matrix contributing to $\arg(\lambda)$. On the other hand, there are practically no new contributions to $\Delta\Gamma$, so that $\Delta\Gamma \ll \Delta M$.

2. *Fourth Quark Generation*

In models of four generations[10], $\Delta M$ gets new contributions from box diagrams with intermediate $W$ and $b'$. For large $|V_{ub'}V_{cb'}|$ and large $m_{b'}$, the new contribution could be of the order of the experimental bounds. There are three CP violating phases in the $4\times 4$ quark mixing matrix contributing to arg($\lambda$). On the other hand, there are practically no new contributions to $\Delta\Gamma$, so that $\Delta\Gamma \ll \Delta M$.

3. *Singlet Left-handed Up Quark*

In models with an $SU(2)$-singlet $u'_L$[11] (or, similarly, $SU(2)$-doublet $u'_R$), there are new contributions to $\Delta M$ from tree diagrams with an intermediate $Z^0$. For large $|U_{uc}|$ ($U$ is here the neutral current mixing matrix) the new contribution could be of the order of the experimental bounds. There are three CP violating phases in the $4 \times 3$ quark mixing matrix contributing to arg($\lambda$). There are new contributions to $\Delta\Gamma$ from $Z$-mediated decays, but the change in $\Delta\Gamma$ is a factor of order one and not orders of magnitude. Consequently, $\Delta\Gamma \ll \Delta M$.

4. *Multiscalar Models with Natural Flavor Conservation*

In multiscalar models with natural flavor conservation[12], $\Delta M$ gets new contributions from box diagrams with intermediate $H^\pm$ and quarks. For very large $\tan\beta \equiv \frac{v_u}{v_d}$ and light charged Higgs, the new contribution could be of the order of the experimental bounds. Diagrams with intermediate $H^\pm$ and $b$ quark could be important; they depend on $(V_{cb}V^*_{ub})^2$ and therefore could contribute (proportionally to the Kobayashi Maskawa phase) to arg($\lambda$). The $D$ decays mediated by $H^\pm$ cannot compete with $W$-mediated decays, so that $\Delta\Gamma$ practically does not change. Consequently, $\Delta\Gamma \ll \Delta M$.

5. *Multi-doublet Models (with Approximate Flavor Symmetries)*

In multiscalar models where flavor-changing scalar couplings are present[13], $\Delta M$ gets new contributions from tree diagrams with intermediate $H^0$. For a flavor changing coupling of order $\sqrt{m_c m_u}/m_W$ and light neutral scalar, the new contribution could be of the order of the experimental bounds. In principle there could be arbitrary new phases in the relevant couplings, though the $\epsilon$ constraint may imply that the phases are small. In this case the contribution to arg($\lambda$) is small.

The $H^0$-mediated decays cannot compete with $W$-mediated $D$ decays, so that $\Delta\Gamma$ practically does not change. Consequently, $\Delta\Gamma \ll \Delta M$.

6. *Leptoquarks*

In models of light scalar leptoquarks[14], $\Delta M$ gets new contributions from box diagrams with intermediate leptoquarks and leptons. For leptoquark couplings $F_{\ell c} F_{\ell u} \geq 10^{-3}$ and leptoquark masses $M_{LQ} \leq 2$ TeV, the new contribution could be of the order of the experimental bounds. There are arbitrary new phases in the $F_{\ell q}$ couplings contributing to $\arg(\lambda)$. The leptoquark-mediated decays practically do not affect $\Delta\Gamma$. Consequently, $\Delta\Gamma \ll \Delta M$.

The conclusions of this brief survey are twofold:

(a) Models of New Physics that contribute significantly to $D^0\overline{D^0}$ mixing can involve new sources of CP violation that are likely to take part in the new mixing amplitudes. One should not assume CP invariance ($\lambda = \overline{\lambda}$) when considering these models. However, *direct* CP invariance ($|A/\overline{A}| = |B/\overline{B}| = 1$) is likely to hold independent of model.

(b) In all reasonable models, the order of magnitude of $\Delta\Gamma$ is similar to the Standard Model. Therefore, the approximation $\Delta\Gamma \ll \Delta M$ (and, consequently, $|p/q| = 1$) is very reasonable.

## 5. Conclusions

We have reviewed the formalism for wrong-sign decays in the limit of $\Delta M \ll \Gamma$, $\Delta\Gamma \ll \Gamma$ and $|\lambda| \ll 1$, leading to the rate equations (2.10) and (2.11). We have shown, contrary to other claims, that the interference terms proportional to $\Delta M t$ cannot necessarily be ignored, even in the case of CP invariance. Furthermore, an examination of relevant New Physics models indicates that CP invariance is likely to be violated if any of these models provides a source of measureable $D^0\overline{D^0}$ mixing. We conclude that an arbitrary interference term should be used in the fits to experimental data. The previous best limit on mixing[4], which came from a time evolution study that neglected interference, is significantly weakened by this argument. Other limits on $D^0\overline{D^0}$ mixing that use methods not dependent on the time evolution are, of course, unaffected by this argument.

## Acknowledgements

One of us (YN) thanks Isi Dunietz and Yuval Grossman for useful discussions. YN is supported in part by the United States – Israel Binational Science Foundation (BSF), by the Israel Commission for Basic Research and by the Minerva Foundation.